# Augmented Rotation-Based Transformation for Privacy-Preserving Data Clustering

Dowon Hong and Abedelaziz Mohaisen

**Multiple rotation-based transformation (MRBT) was introduced recently for mitigating the apriori-knowledge independent component analysis (AK-ICA) attack on rotation-based transformation (RBT), which is used for privacy-preserving data clustering. MRBT is shown to mitigate the AK-ICA attack but at the expense of data utility by not enabling conventional clustering. In this paper, we extend the MRBT scheme and introduce an augmented rotation-based transformation (ARBT) scheme that utilizes linearity of transformation and that both mitigates the AK-ICA attack and enables conventional clustering on data subsets transformed using the MRBT. In order to demonstrate the computational feasibility aspect of ARBT along with RBT and MRBT, we develop a toolkit and use it to empirically compare the different schemes of privacy-preserving data clustering based on data transformation in terms of their overhead and privacy.**

Manuscript received June 9, 2009; revised Dec. 6, 2009; accepted Jan. 4, 2010.
This work was supported by the IT R&D program of MKE, Rep. of Korea (Development of Privacy Enhancing Cryptography on Ubiquitous Computing Environment).

Dowon Hong (phone: +82 42 860 6147, email: dwhong@etri.re.kr) is with the Software Research Laboratory, ETRI, Daejeon, Rep. of Korea.
Abedelaziz Mohaisen (email: mohaisen@cs.umn.edu) was with the Software and Content Research Laboratory, ETRI, Daejeon, Rep. of Korea, and is now with the Department of Computer Science, University of Minnesota Twin Cities, Minneapolis, USA.

## I. Introduction

Data mining has grown to include powerful tools for understanding unknown patterns in huge amounts of data and benefits by drawing ideas from several fields including machine learning, artificial intelligence, pattern recognition, statistics, and database systems [1]. Though the data mining area itself is young compared to other areas, it utilizes smart and powerful algorithms which are adapted from other areas, particularly from the database systems research area. These smart algorithms are essential for understanding data and building models over it, which is a fundamental goal in many business intelligence related areas [2]. For instance, it is possible to improve the quality of services by utilizing patterns of interests using association role mining [3] or data clustering [4], [5]. Also, it is very possible to build predictive models for learning the data by applying the data classification and Bayesian models among other data mining algorithms [6].

While the benefits realized by making data available for the purpose of data mining are great, recent results have shown that privacy is a significant requirement that must be considered along with the data mining results [3]. For ethical and technical reasons, data providers who provide their data for data mining purposes are concerned that their data will not be used for breaching their privacy. For that reason, several privacy-preserving data mining algorithms have been crafted since the initiation of privacy-preserving data mining studies in [3]. These algorithms are designed to provide data mining results over data that conceals or limits access to user identity or data that might lead to user identification. Also, these algorithms have broadened to include most of the known conventional data mining algorithms and consider the mining of data stored at distributed settings for both vertically and horizontally distributed databases while considering several adversarial settings [7].

Recently, privacy-preserving data clustering has been studied for its promising variety of applications in different fields. Privacy-preserving data clustering is a special clustering problem concerned by group data into exclusive sets according to some similarity criterion without breaching the data privacy. To enable this type of clustering, several algorithms have been introduced with specific advantages and disadvantages. For instance, Oliveira and others in [5] and Chen and others in [8] simultaneously introduced the rotation-based transformation (RBT) method in which the data is linearly (and orthogonally) transformed while maintaining a distance-invariance property between data records. This distance-invariance property enables an accurate distance-based clustering over transformed data. Though it is computationally feasible, the RBT was shown by Guo and others in [9] to be vulnerable to the apriori-knowledge independent component analysis (AK-ICA) attack. To mitigate this attack, Mohaisen and others introduced a multiple rotation-based transformation (MRBT) algorithm [4]. Though the MRBT mitigates the AK-ICA attack and allows several distance-based algorithms including special kinds of clustering, it does not enable the conventional clustering on the transformed data [4].

The original contribution of this paper is twofold. First, to enable the conventional clustering on data subsets to be transformed using MRBT where conventional clustering was not previously possible, we introduce an augmented rotation-based transformation (ARBT) scheme which is shown to be efficient for data transformation while mitigating the AK-ICA attack. As a related contribution, we provide an optimization technique which uses the fact that some data is already clustered when applying ARBT and use that for improving the conventional clustering. Second, since both RBT and MRBT are shown theoretically to preserve privacy and to be practical without any empirical evidence or verification, we introduce ppCD, a Java-based toolkit that is designed and used for privacy-preserving data clustering and for performing real measurements for the known clustering algorithms on a typical computing machine. Particularly, ppCD incorporates the RBT, MRBT, and ARBT for privacy-preserving data clustering. Also, it implements a conventional clustering algorithm, namely the *k*-mean clustering algorithm, and an optimized clustering algorithm designed specifically for and benefiting from the application scenario of the ARBT. Among the interesting results realized in this article, we show that transforming 130 MB of storage on a desk requires about 1.6 seconds on a typical computing machine with any of the transformation algorithms we tested in our experiments.

The structure of the rest of this paper is as follows. In section II, we introduce the preliminaries of this work by touching upon the system and data model, data clustering, and data transformation methods for privacy-preserving data clustering. In section III, we introduce our augmented rotation-based transformation (ARBT) scheme for data clustering accompanied by an enhanced version of the *k*-mean clustering algorithm which benefits from the settings of ARBT. In section IV, we evaluate the ARBT in terms of privacy preservation, its security, and overhead. In section V, we introduce our empirical study on the different transformation schemes preceded by developing an evaluation criteria and depiction of the ppCD toolkit. Finally, we draw concluding remarks in section VI.

## II. Preliminaries

In this section, we describe the models used in this article: the system model and the data model. We describe user classification. We define data and, finally, give a description of the rotation-based transformation algorithms.

### 1. Models

*A. System Model*

General data mining systems are designed for mining data according to two different models: the server to server (S2S) model and the client to server (C2S) model [10]. In this work, we consider the C2S model which is depicted in Fig. 1. The C2S model consists of three entities: data providers, warehouse servers, and mining servers [10]. *Data providers* provide data for clustering and are simply the users whose data is of interest to a potential attacker. The *warehouse servers* are storage servers available publicly and accessible by both users and mining servers. Generally, warehouse servers are not trusted, and access to them is not controlled by any authentication or authorization services. Finally, the *mining servers* are used for mining data. Accordingly, mining servers may act maliciously. In the general C2S model, servers do not have direct interaction with users though they gain access to a user's data by requesting it from the warehouse server. In principle, having the warehouse server in the system does not imply any additional security constraints over the scenario of having a miner communicate with users directly. For simplicity, we merge both the mining server and warehouse server in a single miner entity where the mining server initially gains access to data by requesting it from the user.

The ARBT algorithm is a special case. Some interaction takes place between the data miner and the users (clients) to enable conventional clustering. Further details on all possible interactions between users and the mining server concerning the different algorithms are shown in Table 1. In Table 1, $U_A$ and $U_B$ stand for two users (clients) and $M$ stands for the mining server. A single-user system scenario is realized by considering either of the users, and a multi-user system scenario is realized by considering both of them.

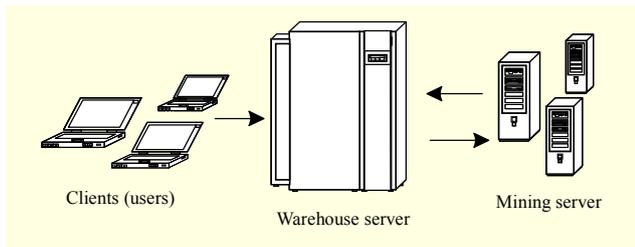

Fig. 1. System model in general data mining application. Note that the directions of arrows here are illustrative, and the notation in Table 1 is considered for each algorithm separately.

Table 1. Communication between users and miner where → and ↔ represent uni- and bi-directional, respectively.

| Algorithm | $U_A, U_B$ | $U_A, M$ | $U_B, M$ |
|---|---|---|---|
| RBT | → | → | → |
| MRBT | ↔ | → | → |
| ARBT | ↔ | ↔ | ↔ |

Table 2. Sample of dataset we use in our implementation.

| Index | A1 | A2 | A3 | A4 | A5 |
|---|---|---|---|---|---|
| 1 | 13.70 | 48.13 | 084.63 | 41.19 | 66.25 |
| 2 | 26.26 | 49.01 | 121.37 | 45.79 | 81.87 |
| 3 | 20.76 | 44.98 | 108.12 | 56.59 | 93.31 |
| 4 | 15.19 | 50.53 | 063.30 | 42.19 | 60.88 |

Table 3. Dataset in Table 1 normalized to unit. Note that, this data is normalized along with larger dataset (50,000 records).

| Index | A1 | A2 | A3 | A4 | A5 |
|---|---|---|---|---|---|
| 1 | 0.3623 | 0.0170 | 0.3239 | 0.1114 | 0.0156 |
| 2 | 0.6450 | 0.0198 | 0.6752 | 0.2340 | 0.0760 |
| 3 | 0.5430 | 0.0072 | 0.5214 | 0.3238 | 0.2179 |
| 4 | 0.1982 | 0.0245 | 0.3656 | 0.0692 | 0.0288 |

*B. User Classification*

Users are either honest or dishonest [11]. Honest users provide their data for mining purposes and do not have any interest in breaching the privacy of other users participating in the protocol by trying to gain access to their private data. Dishonest users may misbehave and seek access to the private data of others by reconstructing it using secret parameters of users participating in the protocol. In this work, we assume all users to have a high incentive to act honestly. Work in [12] suggests the rationale of such assumption.

*C. Data Model*

For data representation, we use the conventional relational database model to store the data [13]. In this model, dataset $A$ consists of $n$ records and $a$ attributes. The index of a record is denoted as $i$ where $1 \leq i \leq n$ and the index of an attribute is denoted as $j$ where $1 \leq j \leq a$. The data in our system is numerical and describes geographical, locational, or financial data. A sample of the data used in our implementation is shown in Table 2, and its normalized image is shown in Table 3. Mathematically, dataset $A$ is represented as $(a \times n)$ matrix.

2. Data Clustering

Given dataset $A$ of $n$ records and $a$ attributes, and given a similarity measure among these records, the data clustering is concerned with dividing the dataset into groups (data subsets) so that the records in one group (named cluster) are more similar to one another, and data records in separate clusters are less similar to one another. The Euclidean distance is a commonly used similarity measure in the continuous data clustering. The Euclidean distance between two records $\check{r}_i = (r_{i1}, \cdots, r_{ia})$ and $\check{r}_j = (r_{j1}, \cdots, r_{ja})$ is $Dist(\check{r}_i, \check{r}_j) = (\sum_{k=1}^{a} (r_{ik} - r_{jk})^2)^{1/2}$.

Some of the known distance-based clustering algorithms that utilize Euclidean distance include the $k$-mean algorithm [14] and $k$-nearest-neighbor ($k$-NN) algorithm [15]. In this article, we use the $k$-mean algorithm for comparing the performance of the different transformation schemes. A slightly modified $k$-mean clustering algorithm is depicted in Fig. 3.

3. Data Transformation Algorithms

In the context of data clustering, both privacy and data mining results are equally important criteria. In order to guarantee both criteria, several methods were introduced considering several potential applications (for example, [16]-[19]); most noticeably, data perturbation. One of these perturbation methods is the RBT in which the data is transformed geometrically while preserving the distance between the data records to enable a distance-based data clustering with high accuracy that reflexes a minimal data loss when performing clustering. Considering several attackers' capabilities, several studies are introduced to test the privacy achieved in the RBT scheme. For instance, it was shown that a combination of reconstruction and distance-based inference attacks can be utilized to greatly breach the privacy of the RBT.

The principle component analysis (PCA) [20] and independent component analysis (ICA) [21], two statistical tools, are shown to be efficient for reconstructing private data transformed using RBT under some operation conditions. In this section, we describe the RBT scheme and the attack on it followed by the MRBT used for mitigating this attack. Also, we provide motivation for our ARBT scheme by showing shortcomings of the MRBT.

*A. RBT*

The RBT is a distance-invariant transformation method. For a dataset $A$ and $R$ represented as an $a \times n$ matrix and an $a \times a$ transformation matrix, respectively, the RBT is mathematically expressed as $Y=RA$. In order to satisfy the distance-invariant property, the transformation matrix $R$ needs to be orthonormal. $R$ is said to be orthonormal if $R=R^{-1}$ (that is, $RR^T=I$, where $I$ is the identity matrix). According to [5], [8], and [20], an orthonormal matrix can be constructed as a square matrix with two non-zero values in each row and column. An example of that is $R = [\check{c}_1, \check{c}_2]$, where $\check{c}_1$ and $\check{c}_2$ are column vectors and defined as $\check{c}_1 = (\cos\theta, -\sin\theta)$ and $\check{c}_2 = (\sin\theta, \cos\theta)$. As an elementary representation, a 2×2 orthonormal matrix can be represented as $R=f(\theta)=[e_{ij}]$, where $e_{ij} = \cos\theta$ iff $i=j$, $e_{ij} = \sin\theta$ iff $i<j$, and $e_{ij} = -\sin\theta$ iff $i>j$. An $a \times a$ orthonormal matrix, where $a$ is an even number, is expressed in terms of a 2×2 orthonormal matrix as a *diagonal block matrix* $R = [R_i]$ where $1 \leq i \leq a/2$ and $R_1 = R_2 = \cdots = R_{a/2}$ [4]. For odd numbers of attributes, the last attribute can be transformed manually with any other previously transformed attribute [5].

RBT is shown to be vulnerable to the AK-ICA attack. The ICA itself is a statistical tool used for signal separation. Given the rotated data and a previously known portion of original private data, with the help of the ICA, the attacker is able to estimate the unknown private data from the rotated data. This attack has been studied in [9] and mitigated in [4]. The mitigation procedure tries to harden the applicability of the AK-ICA by defining several rotation parameters and partitioning the dataset into several data subsets. Then, the data subsets are rotated using rotation matrices generated from the different rotation parameters independently and released for data mining purposes. The detailed procedure of mitigation is shown below.

*B. MRBT*

The MRBT for the mitigating AK-ICA attack uses the long-standing technique of controlling data release to achieve higher privacy. Since the column-wise control of release that provides distance preservation between data columns is useless for data clustering, which is basically performed over rows (records), we use the row-wise control for limiting data release. In the row-wise control, we block each set of records and transform them independently using different transformation parameters. In particular, we group the different records into different groups and rotate them using different rotation matrices generated by different random instances according to the previously mentioned method. This control of transformation and data release preserves full distance over records for parts of the columns. In particular, this distance enables correct clustering over the subsets. MRBT is performed as follows [4].

1. The data owner normalizes the data to the unit.
2. According to some parameter $m$, the data owner divides the data into $m$ equal parts defined as $A= \{A_1, \cdots, A_m\}$. $A$ is expressed as the block matrix $A= [A_i]$: $1 \leq i \leq m$.
3. The data owner generates $m$ different random seeds ($s_1, \cdots, s_m$). Using each seed $s_i$, the data owner generates an orthogonal matrix $R_i=f(s_i)$ (as detailed in section II.3.A) for transforming the corresponding sub-matrix of $A$.
4. The data owner transforms his data as $Y=[Y_i]=[R_iA_i]$, where $1 \leq i \leq m$ and releases $Y$ for data clustering purpose.

The resulting rotation preserves the inner product between the corresponding records in the original data. Also, it preserves the inner product between two records falling into the same corresponding subsets. However, the inner product is not preserved for records other than those mentioned here. The first part can be easily proven to be correct given that these records are transformed using the same matrix. Similarly, we prove the second part as follows. Let $Y_A=[Y_{Ai}]=[R_iA_i]$ and $Y_B=[Y_{Bj}]=[R_jB_j]$. The inner product between $Y_A$ and $Y_B$ is $Y_A^T Y_B= [Y_{Ai}^T Y_{Bj}]$, where $1 \leq i \leq m$ and $1 \leq j \leq m$. For the diagonal part of this product matrix ($i=j$), it is easy to verify the preservation of the inner product since $Y_{Ai}^T Y_{Bj} = (R_iA_i)^T R_iB_i = A_i^T R_i^T R_iB_i = A_i^T IB_i = A_i^T B_i$.

In spite of mitigating the AK-ICA attack, which is a great merit [4], the MRBT does not allow conventional clustering on data that belongs to two different subsets. To enable that, we introduce an ARBT for conventional clustering of transformed data and its privacy.

### III. ARBT

In order to overcome the shortcoming of the MRBT, we introduce the ARBT scheme that takes both conventional clustering and privacy into account. The ARBT consists of two parts applied separately on the side of the client (data owner) and the server. Before detailing both parts, we motivate for the ARBT by introducing the linearity property of transformation.

1. Motivation

The linearity of the RBT is an interesting property that can be used for enabling conventional clustering over data transformed using the MRBT. Let $R_1 = f(\theta_1)$, $R_2 = f(\theta_2)$, and $R_3 = f(\theta_3)$. We state that $R_1 R_2 = R_3$ if $\theta_1 + \theta_2 = \theta_3$, where $f$ is a function that constructs an orthogonal matrix from a random seed $\theta$ according to the procedure explained earlier. Proving this statement for 2×2 transformation matrices is direct and can be easily generalized to an orthonormal transformation matrix of any size. Let $R_1 = [\check{r}_{11}, \check{r}_{12}]$ and $R_2 = [\check{c}_{21}, \check{c}_{22}]$, where $\check{r}_{11} = (\cos\theta_1, \sin\theta_1)$ and $\check{r}_{12} = (-\sin\theta_1, \cos\theta_1)$ are row vectors of $R_1$, and $\check{c}_{21} = (\cos\theta_2, -\sin\theta_2)$ and $\check{c}_{22} = (\sin\theta_2, \cos\theta_2)$ are column vectors of $R_2$. The product, $R_3 = R_1 R_2 = [\check{r}_{31}, \check{r}_{32}]$, is $\check{r}_i \check{r}_j = \cos(\theta_1+\theta_2) = \cos\theta_3$ if $i = j$, $\check{r}_i \check{r}_j = \sin(\theta_1+\theta_2) = \sin\theta_3$ if $i<j$ and $\check{r}_i \check{r}_j = -\sin(\theta_1+\theta_2) = -\sin\theta_3$ when $i>j$, from which we conclude the soundness of the claim.

Since any diagonal block matrix such as the one used in ARBT is also orthonormal, the above statement can be generalized to transformation of any size. Given this linearity property, now we explain the ARBT by discussing the procedure performed at the client and server sides, respectively.

Given the linearity property, our goal can be achieved in a straightforward manner: if we are given two datasets that are transformed using two different parameters, we can make them look as though they are transformed using a single parameter, and hence maintain distance invariance property over their records using the unification property that utilizes linearity.

## 2. Client Side

The procedure of the ARBT at the client side consists of three phases. Two of these phases are performed for rotating the data initially and are typically the same as the procedure performed in the MRBT scheme. On the other hand, the third phase is performed when clustering results are to be computed over two data subsets using the conventional clustering method. In the description below, we consider a single user model though it should be clear that extension to a multi-user model is straightforward as shown in [4]. The three different phases of the client side are detailed in Fig. 2.

## 3. Server Side

The data miner follows the same procedure as in MRBT in response to steps 1 and 2 of the client side: the server will be able to learn clusters over each data subset separately [4]. However, after the client performs step 3 and releases the parameter $\theta_{ij}$ upon the server's request, the server performs the following. First, the server computes a $a \times a$ orthonormal matrix $R_{ij}$. Then, assuming that $A_i$ is the least transformed data set, the server computes $Y_i^* = R_{ij} Y_i$, constructs the block matrix $Y = [Y_i^*, Y_j]$, and learns the clusters over $Y$. Note that the parameter

1. Initialization: Data is divided and transformation parameters are generated as follows:
   a. Data owner with dataset $A$ as $a \times n$ matrix divides $A$ vertically into $m$ block-matrices, where each has $c = (n/m)$ records (where $c > a$). $A$ is then notated as $A = [A_i]$.
   b. The data owner generates $m$ different seeds $s_1, \cdots, s_m$ and computes $R_1 = f(s_1), \cdots, R_m = f(s_m)$, where $R_i$ is an $a \times a$ orthonormal matrix. Each $R_i$ is associated with $A_i$ with the same index.
2. Data rotation and release: Each block-matrix is transformed independently and released to the miner. That is, the data owner computes and publishes $Y_A = [Y_{Ai}]$, where $Y_{Ai} = R_i A_i$. Note that mining is only possible on records in each subset (as in MRBT scheme [4]).
3. Further data release: Upon request, the data owner computes and releases parameters that make clustering on records that belong to two different subsets possible. For two datasets, $A_i$ and $A_j$, transformed using $R_i$ and $R_j$, which are defined as $R_i = f(\theta_i)$ and $R_j = f(\theta_j)$, the user computes and releases $\theta_{ij}$

$$\theta_{ij} = \begin{cases} \theta_j - \theta_i, & \theta_j > \theta_i, \\ 360 - (\theta_i - \theta_j), & \theta_i > \theta_j. \end{cases}$$

Fig. 2. Description of the client side procedure in ARBT.

This algorithm takes $Y_i^*$ and $Y_j$ as two clustered subsets and computes a single set of clusters over their merged dataset. Recall that the two subsets are unified according Fig. 2.
1. For each cluster $C$ in data subset $Y_i^*$ with a centroid $C_e$
   a. Compute the distance between $C_e$ and the centroid of different clusters in the subset $Y_j$.
   b. Add the data records of the cluster $C$ to the cluster in $Y_j$ that has the closest direct (that is, Euclidian) distance to $C_e$.
2. Lloyd algorithm: For each record in each cluster in the merged clusters set in step 1:
   a. Compute the distance between the record in question and the centroid of each cluster in the merged set of clusters.
   b. Attach the record to the cluster that has the closest direct distance to it (with respect to the cluster's centroid).
   c. Re-compute the centroid of updated cluster as the average of data records in that cluster.
   d. Repeat step 2 until no (or almost no) records move out of their current cluster.

Fig. 3. Description of the modified $k$-mean clustering algorithm used to cluster two subsets transformed by ARBT.

$\theta_{ij}$ always unifies the transformation of data subset $A_i$ to $A_j$ in a clock-wise direction.

Now, in order to exploit the fact that both of the data subsets are already clustered, we introduce a method that reduces the overall clustering overhead in terms of computation in our ARBT scheme. To achieve that, we introduce a modified clustering algorithm by assuming that the number of clusters to be learned from the whole dataset which results from merging the two subsets is same as the number of clusters computed over each of the two subsets separately.

In our modified clustering algorithm, we consider one of the two subsets as an initial set of clusters (either $Y_i$, after its transformation into $Y_i^*$ or $Y_j$ itself). After that, we add the different records into one among the different clusters according to their distance to that cluster's centroid. Then, a Lloyd procedure is performed for stabilizing the final clusters. The procedure of the modified algorithm is in Fig. 3.

Note that step 2d considers a stop criterion of iterations which is realized after a constant number of iterations or real monitoring of the change in each cluster.

## IV. Theoretical Evaluation of ARBT

In this section, we evaluate the ARBT. We consider both privacy and security of the scheme under possible threats. For privacy, we consider how much benefit the ARBT exposes for an attacker by revealing larger subsets of data that enables higher accuracy when applying AK-ICA. For security, we study how linear regression can be used for reconstructing original transformation parameters and point out the number of operation times at which the ARBT is considered safe. We further evaluate the resource requirements of ARBT in terms of computation, communication, and additional memory, if any.

### 1. Privacy Evaluation

The privacy achieved in the MRBT basically depends on the number of data subsets, $m$, according to which the whole dataset is divided. In ARBT, in order to enable conventional clustering, we merge different subsets as if they are transformed using a single transformation parameter. Therefore, the ARBT directly reduces $m$, which is an essential parameter for the degree of mitigation of the AK-ICA attack. That is, reducing $m$ will directly reduce the mitigation of the AK-ICA attack achieved in MRBT. As pointed out in [4], both AK-ICA and MRBT are data-driven algorithms. For instance, the AK-ICA attack may work efficiently on data with Gaussian distribution while it works less efficiently on data with geometrical distribution. Also, the MRBT can mitigate the AK-ICA attack greatly with smaller $m$ for some data distributions while other distributions require larger $m$ at the expense of data utility [4]. Therefore, given a mitigation level of the AK-ICA to be achieved in the ARBT on *a specific dataset*, we can certainly compute and experimentally verify the minimum $m$ and the maximum allowed number of data subsets to be merged in a single subset at average. This aspect is verified in our experimentation on the Banker dataset used in [4].

### 2. Security Evaluation

To illustrate the security issue related to ARBT, consider the following example. Let $A_1$, $A_2$, and $A_3$ be three different datasets and their transformed images be $Y_1$, $Y_2$, and $Y_3$. Assume that the released parameters for further transformation in ARBT are $\theta_{12}$, $\theta_{13}$, and $\theta_{23}$ which are released for computing $R_{12}$, $R_{13}$, and $R_{23}$, respectively. Using these parameters, we can compute the following linear system of equations (a) $\theta_{12} = \theta_2 - \theta_1$, (b) $\theta_{13} = \theta_3 - \theta_1$, and (c) $\theta_{23} = \theta_3 - \theta_2$. One can easily check the solvability of this system by observing that the system consists of three linearly independent equations in three variables from which the attacker can break the security of the original RBT. To prevent that, we limit the released parameters so that they can not be used for recovering the original transformation parameters. For instance, releasing $m-1$ parameters that construct $m-1$ linearly independent equations is considered safe.

Note that the transformation to unify more than two datasets can be safely performed as well since it is not necessarily required to release parameters to unify all of the $m-1$ pairs of two basic data subsets unification. Also, note that the ARBT is basically designed for data of limited use under the assumption that data collected for clustering purpose will not be of interest once clusters are learned from it. However, we can strengthen the ARBT for permanent data by reapplying the procedure in Fig. 2 periodically and changing the initial parameters so their revelation will not affect the scheme.

Though our scheme does not provide any guarantee against colluding mining servers since we assume a single miner motivated by our application settings, users can wisely release parameters so that colluding servers don't breach their privacy through these parameters. If users respond with unification parameters for each requesting server, the utility of our scheme will degrade linearly in relation with the number of miners at worst. On the other hand, if the user limits the set of parameters that she would like to release, the utility can be maintained as high as in the single server model.

Note that the security analyzed as per the example in this section implies two party settings (single user and single miner) while three-party settings can be either with two miners and one user (as in the case above) or two users and one miner. The later case of three-party settings is similar to the two party case since we assume honest and trusted users. Though this assumption is not the optimal desirable form of separation between users, it has been used and advocated for applications in literature such as the one in [6].

### 3. Overhead Evaluation

While the memory required in the ARBT scheme is the same as in RBT and MRBT, additional computation is required for learning clusters over unified data subsets, and communication is required for exchanging additional parameters. Since the client generates an additional parameter, $\theta_{ij}$, and passes it to the mining server to compute $R_{ij}$ communication required for transferring, this parameter is constant per single ARBT operation. Computation required for computing clusters over the unified data is shown in section V where we suggest methods to reduce the overhead required in the naive scenario with MRBT. The extra overhead required for generating an orthonormal matrix of unification on the server is negligible compared to the processing required for clustering tasks.

While in ARBT one of the two subsets is transformed in the further release step, both subsets are transformed in the naive scenario using a single transformation parameter. At the server side, while the ARBT reduces the overhead of clustering by considering previously clustered subsets as initial clusters, the naive scenario necessitates clustering with initially empty clusters with more overhead. Further details on this note are shown in experiment 10.

## V. Empirical Study

Here we introduce an empirical study on privacy-preserving data clustering. Though the overhead of the client is probably the most essential element in the context of PPDM feasibility and applicability, this study considers the overhead consumed of the client and server to compare the different schemes fairly.

### 1. Evaluation Criteria

In this study, we develop two criteria for evaluating the different schemes. These criteria are resources consumption (that is, overhead) and the achieved privacy. While the overhead is mainly expressed as time of computation, the achieved privacy is evaluated based on the degree of mitigation to the AK-ICA attack. The two criteria used are detailed as follows.

#### A. Overhead

We study the overhead required on the client and the server sides. Concerning the client, we evaluate the overall computation overhead required for data transformation in the different schemes. We also consider the additional overhead required for both the MRBT and ARBT as they differ from the conventional RBT scheme. On the server side, we evaluate the computation overhead required for clustering different sets of data. We also study the impact of initial centroids' selection on the required computation overhead in term of iterations and time. All measurements for the overhead evaluation are computed using the privacy-preserving data clustering (ppCD) toolkit.

#### B. Privacy

We study the privacy achieved using the different transformation schemes based on their mitigation to the AK-ICA attack. For ARBT, since the transformation part of ARBT is performed at the server side, we study both privacy and overhead for different percents of pairs of subsets that are transformed using the ARBT in order to enable conventional clustering.

### 2. Overview of ppCD

The Java-based ppCD toolkit implements RBT, ARBT, and MRBT. It also implements several normalization procedures such as unary-norm, $z$-norm, and min-max norm. It also incorporates different clustering methods, including the $k$-mean clustering algorithm, $k$-nearest-neighbor clustering algorithm, and a modified $k$-mean clustering algorithm which is designed specifically for the ARBT scheme. The functional client of ppCD, depicted in Fig. 4, takes raw data, normalizes it using the user's inputs for a normalization procedure, enables users to input transformation parameters (or assign them at random), and transforms data according to the selected algorithm.

The functional server enables the administrator to load transformed data for processing, select a clustering algorithm, supply parameters for the clustering algorithm, and compute the clusters over the transformed data using the selected algorithm. The server also provides the capabilities of computing statistical properties on the transformed data and save them along with the computed clusters. A functional description of the server side is depicted in Fig. 5.

The data used in our implementation and experiments includes real and synthetic data. The real data, which was also used in [4], is the Banker dataset which consists of 50,000 records. Our synthesis data consists of $10^6$ records and was generated using a random generator with specific distribution (details are in experiment 4). To trace the precise impact of data size on the different criteria in the different schemes, we divide the Banker dataset incrementally as shown in Table 4.

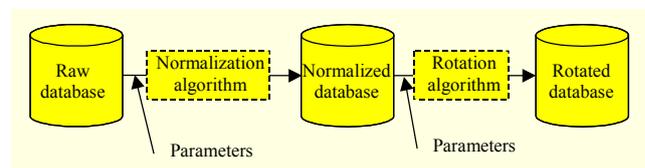

Fig. 4. Client module of the ppCD.

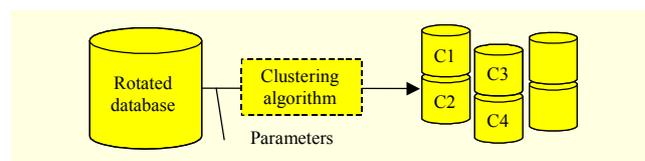

Fig. 5. Server module of the ppCD.

Table 4. Datasets. 'Size' stands for the theoretically computed size as double data type, 'Size (d)' stands for the size on the desk, and 'time' stands for the time required for rotating the corresponding dataset on the experimentation machine.

| Set | Record | Attribute | Type | Size | Size (d) | Time (ms) |
|---|---|---|---|---|---|---|
| $s_1$ | 3,125 | 4 | num | 100 | 170 | 2.147 |
| $s_2$ | 6,250 | 4 | num | 200 | 338 | 4.088 |
| $s_3$ | 9,375 | 4 | num | 300 | 507 | 5.037 |
| $s_4$ | 12,500 | 4 | num | 400 | 677 | 8.005 |
| $s_5$ | 15,625 | 4 | num | 500 | 847 | 10.060 |
| $s_6$ | 18,750 | 4 | num | 600 | 1,016 | 12.060 |
| $s_7$ | 21,875 | 4 | num | 700 | 1,186 | 14.041 |
| $s_8$ | 25,000 | 4 | num | 800 | 1,356 | 15.963 |
| $s_9$ | 28,125 | 4 | num | 900 | 1,525 | 18.025 |
| $s_{10}$ | 31,250 | 4 | num | 1,000 | 1,697 | 20.029 |
| $s_{11}$ | 34,357 | 4 | num | 1,100 | 1,868 | 21.814 |
| $s_{12}$ | 37,500 | 4 | num | 1,200 | 2,035 | 23.871 |
| $s_{13}$ | 40,625 | 4 | num | 1,300 | 2,205 | 25.822 |
| $s_{14}$ | 43,750 | 4 | num | 1,400 | 2,376 | 27.974 |
| $s_{15}$ | 46,875 | 4 | num | 1,500 | 2,546 | 30.220 |

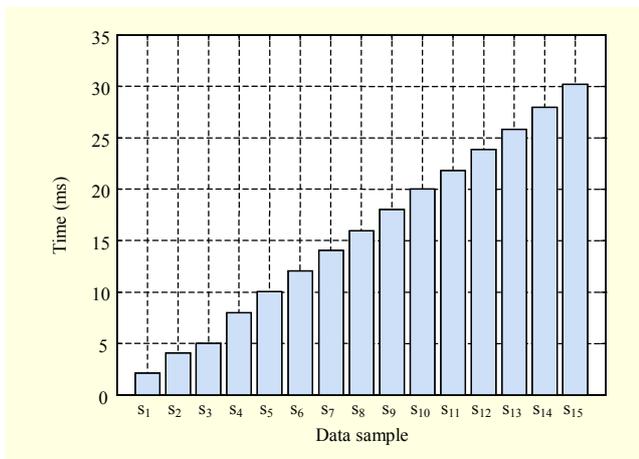

Fig. 6. Mean time required for transforming different datasets. The time linearly depends on the number of records.

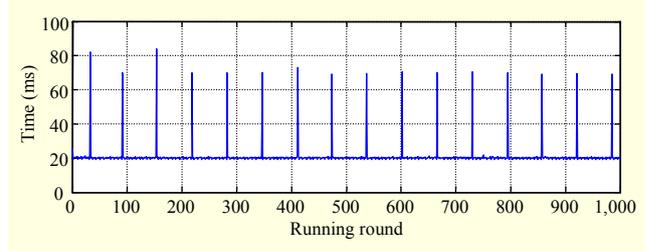

Fig. 7. Raw time measurements of time required for MRBT on dataset $s_{10}$ for 500 times (rounds).

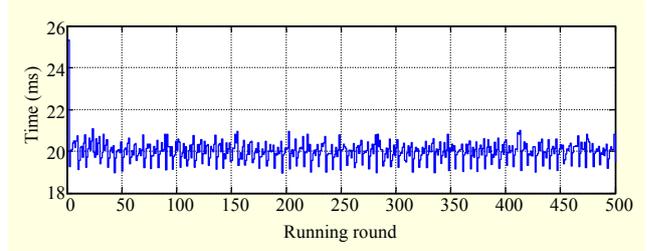

Fig. 8. Filtered time measurements of time required for MRBT on $s_{10}$ applied for 500 times.

3. Empirical Study

To study the feasibility of the different RBT schemes empirically, we use our ppCD toolkit. In our experiments, we consider the overhead required by the client and server alike. We compare the different schemes in term of their resource consumption and privacy achieved as detailed in the following experiments. Note that all of the experiments are performed on a computing machine equipped with an Intel Core 2 Quad CPU that utilizes a 32-bit data bus and operates at 2.5 GHz with 3.25 GB of RAM.

**Experiment 1.** In this experiment, we measure the average time required for transforming different datasets according to the RBT scheme in section II.3.A. For different datasets (shown in Table 4), we execute RBT and measure the required computation. As expected, we found that, on average, the required computation time linearly depends on the number of records as shown in Table 4 and rendered in Fig. 6.

**Experiment 2.** In this experiment, we tried to maximize the accuracy of the time measurement in the MRBT case. Because the ppCD toolkit shares the computing machine's resources with other running processes, measured time in an experiment may not be as accurate as needed. To eliminate the error in measurement that results from this scenario, we run the MRBT on $s_{10}$ 500 times and measure the execution time (shown in Fig. 7). We observed that some of the measured times are greatly higher than the majority of measurements. For higher accuracy, we filtered these ambiguous measurements and replaced them with the average measurements (shown in Fig. 8). The average time required for processing is then computed over all the measurements, including the filtered measurements, which greatly matches with results in experiment 3.

**Experiment 3.** We evaluate additional computation overhead represented as time and required for the MRBT over the RBT. As shown in Fig. 9, for a dataset of 31,250 records transformed using the MRBT at $m=100$, we found that the additional computation required in MRBT is slight compared to the initial overhead required in RBT. While RBT requires 19.8910 ms, the MRBT requires 20.7804 ms for the same dataset. The extra time in MRBT over the time required in RBT is 0.8894 ms, which is 4.47% of the overall overhead. Though this overhead is small in relation with merit realized in MRBT, it is even smaller for larger datasets.

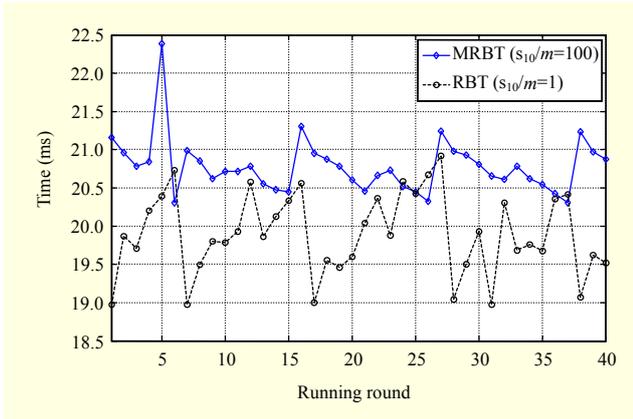

Fig. 9. Running time of the MRBT vs. RBT on the client side (transformation only) for $s_{10}$. The average times required for MRBT and RBT are 20.78 ms and 19.8910 ms, respectively.

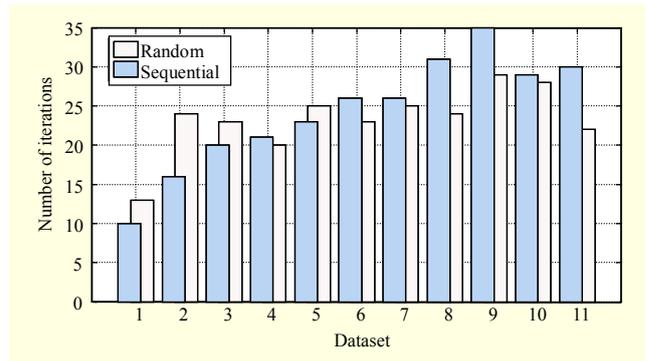

Fig. 10. Number of iterations for achieving stabilized clusters for random vs. sequential selection of initial centroids.

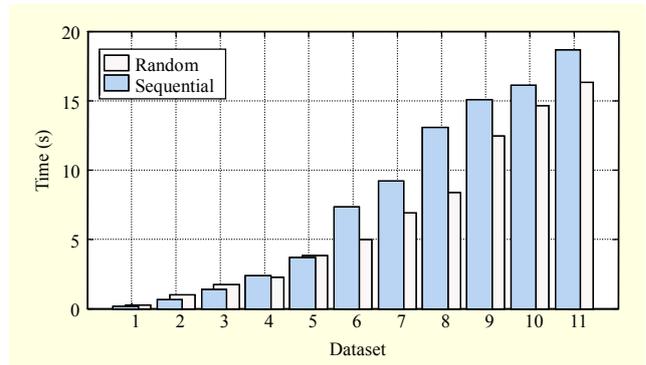

Fig. 11. Time required for achieving stabilized clusters.

**Experiment 4.** In this experiment, we generated normally distributed synthesis data which has a statistical mean equal to its variance ($\mu=\sigma^2=100$). The dataset has $10^6$ records and each record has 10 attributes (about 130 MB on a desk). We transformed the dataset using the RBT to measure its feasibility. We realized that transforming the whole dataset takes about 1.601 seconds. In the same dataset, MRBT with $m=100$, took 1.614 seconds, which is about 0.88% additional overhead.

**Experiment 5.** We observed that required computation overhead for transforming data in the ARBT, as shown in steps 1 and 2 of Fig. 2, are equal to the overhead required in MRBT. Furthermore, we observed that the computational overheard required for step 3 in Fig. 2 is equal to that required in the RBT scheme for half of the dataset over which conventional clustering is to be performed. For instance, if a dataset is divided into two parts, then transformed using ARBT and re-transformed to enable conventional cluster, the overall computation overhead required at the client side is one and a half times of the overheard required in the RBT or the MRBT (since both schemes require almost the same amount of overhead). However, compared to the naive scenario described in section IV.3, ARBT requires only half of the overhead required for further data transformation.

**Experiment 6.** We learned the impact of initial centroids on the number of required iterations for the Lloyd algorithm used for computing final clusters. We studied the case of random against sequential centroid assignment. Figure 10 shows the number of iterations required for data subsets in Table 4. We realized that though the random selection of initial centroids does not necessarily reduce the number of iterations as shown for small data subsets, it reduces the overhead when the number of records to be clustered is large. For instance, the average number of iterations required in the random scenario is 23.2727 iterations per dataset, while it is 24.2727 per dataset for sequential scenario with the final clusters $k=7$.

**Experiment 7.** The previous experiment was performed again to measure the time required for clustering datasets in both of the random and sequential assignment scenarios. We realized that the average time required for clustering a set for the random scenario is 6,620.4 ms, while it is 7,980 ms for the sequential assignment scenario. A comparison between the two scenarios for the different datasets is shown in Fig. 11.

**Experiment 8** (AK-ICA on ARBT-1). We studied the impact of the AK-ICA attack on ARBT. Particularly, we considered the scenario where several percents of all possible data subset pairs (that is, $m–1$ for safety) are transformed to enable the conventional clustering and measured the mitigation of the AK-ICA attack on the resulting dataset for fixed percent of known private data to an attacker. In this experiment, we considered the whole banker dataset and $m=100$. We also set initial percents of known private data to the attacker (as 5% and 10%) and studied the impact of AK-ICA on ARBT when different portions of data are transformed. The results of this experiment are shown in Table 5. Note that MRBT indicates 0% of the subsets are transformed using ARBT. Also, 100% for all two-subset pairs (that is, $m–1$) in ARBT is equivalent to the case of MRBT at $m=50$ [4]. Though the error of the attacker's estimation of the private data is relatively lower than that achieved by MRBT, a degree of mitigation of the AK-ICA attack is possible when applying ARBT.

Table 5. Impact of AK-ICA attack on ARBT for different percents of private data known to attacker and transformed subsets (initial $m = 100$ and final $m = 50$).

| Known percent | MRBT | 25% | 50% | 75% | 100% |
|---|---|---|---|---|---|
| Mitigation (5%) | 0.970 | 0.835 | 0.724 | 0.590 | 0.251 |
| Mitigation (10%) | 0.963 | 0.817 | 0.698 | 0.531 | 0.220 |

Table 6. Impact of AK-ICA attack on ARBT for different percents of private data known to attacker and transformed subsets (initial $m = 200$ and final $m = 100$).

| Known percent | MRBT | 25% | 50% | 75% | 100% |
|---|---|---|---|---|---|
| Mitigation (5%) | 0.981 | 0.978 | 0.977 | 0.973 | 0.972 |
| Mitigation (10%) | 0.974 | 0.971 | 0.970 | 0.968 | 0.965 |

**Experiment 9** (AK-ICA on ARBT-2). We repeated the previous experiment with different initial parameters of MRBT. We initially set $m=200$ and performed the same experiment for the same percents of known private data to the attacker against the same percents of transformed pairs of data subsets using the ARBT. The mitigation degree of the AK-ICA attack is shown in Table 6, where the final value of $m$ is 100 if all pairs of subsets ($m–1$) are transformed using ARBT. From this experiment we conclude that ARBT provides high mitigation of the AK-ICA attack for carefully assigned parameters.

**Experiment 10** (Optimization of ARBT). To test the optimization scenario described in Fig. 3, we performed this experiment and compared the result to the naive scenario in which the user transforms both of the original two subsets using a single parameter at the same time and releases them to the miner. On the miner side, the whole procedure of the

$k$-mean clustering is performed. In this experiment, we realized that clustering $s_{10}$ in our ARBT scheme can save up to 32% of the computation overhead. More precisely, while the naive scenario takes 14.7 seconds for clustering $s_{10}$, the optimization scenario for RBT takes only 9.96 seconds.

## VI. Conclusion

In this article, we introduced the ARBT that enables conventional clustering over data transformed using MRBT. To improve its applicability for data clustering, we introduced an enhanced clustering algorithm that considers the scenarios of ARBT deployment where some of the data transformed using ARBT is already clustered. Unlike RBT and MRBT, the ARBT scheme both mitigates the AK-ICA attack and enables conventional data clustering.

To show the feasibility of the different transformation schemes, we introduced an extensive experimental study using the ppCD toolkit. This study concluded that the overhead required for mitigating the AK-ICA attack, in both of ARBT and MRBT, is almost negligible on the server side. Also, it showed that all transformation schemes are very feasible on typical computing machines even for large datasets.

Since this study considered only empirical measurements for the impact of the AK-ICA, which is a necessary contribution for understanding the behavior of the different transformation schemes, in the near future we will investigate the development of mathematical framework that expresses the relationship between the AK-ICA attack, its mitigation level, and the different parameters of the ARBT.

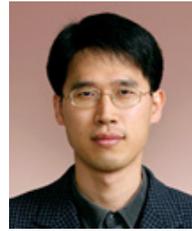

**Dowon Hong** received his BS, MS, and PhD degrees in mathematics from Korea University, Seoul, Korea, in 1994, 1996, and 2000, respectively. He is currently a principal member of the engineering staff and team leader of the Cryptography Research team at ETRI, Korea, where his research interests are broadly in the area of applied cryptography, networks security, and digital forensics.

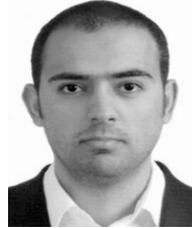

**Abedelaziz Mohaisen** is a PhD student at the University of Minnesota, USA. He was a member of the engineering staff at ETRI, Korea, from 2007 to 2009. He received his BE degree in computer engineering from the University of Gaza, Palestine, in 2005, and the ME degree in information and telecommunication engineering from Inha University, Korea, in 2007. His research interests include networks security, data privacy, and cryptography. He is a member of ACM, IEEE, and KSII.